\documentclass{article}

\pdfoutput=1

\usepackage{fullpage}
\usepackage[space]{cite}
\usepackage{subfigure}
\usepackage{amsfonts,amssymb,amsmath,citesort}
\usepackage{theorem}
\usepackage{color}
\usepackage{colortbl}
\usepackage{multirow}
\usepackage{tikz}
\usepackage[linesnumbered, ruled, vlined]{algorithm2e}
\usepackage{rotating}
\usepackage{threeparttable}
\usetikzlibrary{arrows}
\usepackage{cases}

\newtheorem{theorem}{Theorem}

\newcommand{\remove}[1]{}
\theoremstyle{definition}

\newcommand{\bpr}{\noindent{\em {Proof\/.} }}
\newcommand{\bopr}{\noindent{\em {Outline of the Proof\/.} }}
\newcommand{\epr}{\hspace*{\fill}$\Box$\medskip}
\newtheorem{lemma}{Lemma}

\usepackage{pgf}
\usepackage{tikz}
\usetikzlibrary {arrows}

\colorlet{rouge}{red!80!black}
\colorlet{bleu}{blue!80!black}
\colorlet{vert}{green!80!black}
\colorlet{jaune}{yellow!80!black}

\newcommand{\toolarc}[2]{(#1\rightarrow#2)}

\author{Jacob Chakareski\footnote{Signal Processing Laboratory - EPFL,
    Switzerland}\ \ \ Pascal Frossard\footnotemark[1] \ \ \ Herv\'e
  Kerivin\footnote{Department of Mathematical Sciences - Clemson
    University, USA}\ \ \ Jimmy Leblet\footnote{Department Informatique -
    Institut Telecom~;~Telecom Bretagne,France}\ \ \ Gwendal
  Simon\footnotemark[2]}

\title{A note on the data-driven capacity of P2P networks}

\date{}

\begin{document}

\maketitle

\begin{abstract}

  We consider two capacity problems in P2P networks. In the first one, the nodes have an infinite amount of data to send and the goal is to optimally allocate their uplink bandwidths such that the demands of every peer in terms of receiving data rate are met. We solve this problem through a mapping from a node-weighted graph featuring two labels per node to a max flow problem on an edge-weighted bipartite graph. In the second problem under consideration, the resource allocation is driven by the availability of the data resource that the peers are interested in sharing. That is a node cannot allocate its uplink resources unless it has data to transmit first. The problem of uplink bandwidth allocation is then equivalent to constructing a set of directed trees in the overlay such that the number of nodes receiving the data is maximized while the uplink capacities of the peers are not exceeded. We show that the problem is NP-complete, and provide a linear programming decomposition decoupling it into a master problem and multiple slave subproblems that can be resolved in polynomial time. We also design a heuristic algorithm in order to compute a suboptimal solution in a reasonable time. This algorithm requires only a local knowledge from nodes, so it should support distributed implementations.

  We analyze both problems through a series of simulation experiments featuring different network sizes and network densities. On large networks, we compare our heuristic and its variants with a genetic algorithm and show that our heuristic computes the better resource allocation. On smaller networks, we contrast these performances to that of the exact algorithm and show that resource allocation fulfilling a large part of the peer can be found, even for hard configuration where no resources are in excess.


\end{abstract}

\section{Introduction}

Distributed architectures offer cost effective solutions to the
deployment of large scale data delivery services. Peer-to-peer
solutions have received a lot of interest from the research community
and recently also from the industry. Typically, they permit to share
resources among the different peers in order to offer an adequate
quality of service to all the actors of the system. We can distinguish
two types of resources in distributed systems. Owing to economics
terminology, we denote as \emph{rival} the resources that cannot be
simultaneously allocated to multiple users~\cite{benkler}. In computer
communications, the storage capacity or the uplink bandwidth are
typically rival resources. Other resources are called
\emph{non-rival}.

Peer-to-peer architectures are appealing since the total amount of
available rival resources increases with the number of clients in
absence of selfish behavior. This provides improved scalability
compared to centralized solutions. However, the problem of resource
management in peer-to-peer systems is still very challenging. First,
peers can only allocate resources (\textit{i.e.}, reserve upload
bandwidth) to the peers they know, so it is possible that all
neighbors of a given peer cannot satisfy its demand, although
resources are in excess in another location in the overlay. Studying
the capacity of overlay networks is emerging as an important related
subject~\cite{cui2004aoc,4378424,streamingcapacity,Yang2006175}. Second,
the circulation of non-rival resources (\textit{i.e.}, data) has an
impact on the allocation of rival resources. For instance, in a live
streaming system, a peer may have no fresh data to send to one of its
neighbors, so the upload bandwidth allocated to this neighbor will be
unused. Efficient large-scale content distribution is another major
area of related research~\cite{MaTwGk07,massoulie2008ros}.

In this paper, we address the problem of resource allocation from an
optimization standpoint. Each peer\footnote{Client, node, vertex or
peer are used interchangeably in the document.} is characterized by
its \emph{capacity}, the amount of rival resources it is able to
allocate to other peers. In many cases, the capacity of a peer is its
upload bandwidth, but it can also represent the storage capacity in
distributed back-up services, or the processing power in grid
computing. In parallel, each peer is also characterized by its
\emph{demand} that represents the minimal amount of resources the
system should allocate to it, as otherwise the peer would quit the
system. The demand can be a parameter of the system (e.g., the video
bitrate of the content in live streaming systems) or the individual
need of a node.


We consider that the network overlay is given. In such a model, a peer
can only allocate its resources to its direct neighbors in the
topology, this set of neighbors being fixed.  This is the case when
the overlay is used for several purposes, for example in P2P virtual
worlds, the overlay for event notification is also used for
multimedia. This is also the case when the overlay construction is
driven by external guidelines, for example network locality, peers
that are close in the network should be preferentially connected in
the overlay.

Contrarily to most prior work, we do not consider that the network
links have limited capacities but rather that the nodes have a limit
in the resource they could contribute to their neighbors. This
corresponds to recent models where it has been shown that the capacity
bottlenecks are not located in the backbone but rather at the edges of
the network in the current
Internet~\cite{streamingcapacity,MaTwGk07}. The challenge in the
resource management problem is therefore to be able to match the
demands of the peers with the constrained capacities of their
neighbors.

We study in this paper two instances of the problem of the resource
allocation and propose a theoretical groundwork on the topic of
peer-to-peer capacity. We first compute the capacity of the
peer-to-peer system in the stationary regime in a problem similar to
the performance analysis of bit-torrent systems~\cite{QiuS04}. We
neglect the non-rival resources and consider that peers have enough
data to fully use the rival resources that have been allocated to
their neighbours. We show that maximal resource allocation can be
computed in polynomial time by reducing the problem to the computation
of a maximal flow in a bipartite graph.

We then relax the assumption on the availability of non-rival
resources, and we consider that the capacity of the system is
dependent on the availability of data in the nodes. This second
resource allocation problem is able to consider the dynamics of the
system as in the example of a source broadcasting a non-rival
resource. A node can allocate its resources only if its demand is
fulfilled first. It leads to a multi-constrained optimization problem
whose objective is to maximize the overall quality of service among
the fulfilled nodes, or equivalently to determine the \emph{maximum
number of peers whose demand is fulfilled}. We show that this problem
is however NP-complete. We present a promising Benders'
decomposition~\cite{benders1962pps} of this optimization problem into
one master problem and up to $n-1$ sub-problems, with $n$ being the
number of nodes. We then show that the subproblems can be solved in
polynomial-time, which is promising for the design of fast solution
techniques. We also propose heuristic-based algorithms to the resource
allocation problem, which offer suboptimal yet practical solutions for
large-scale distributed systems. We finally analyze the performance of
the proposed algorithms for networks of small and medium scales.

\section{Overlay Resource Allocation}


\subsection{Framework}

We model the overlay as an undirected graph $G=(V,E)$ where an edge
between two nodes $u$ and $v$ in the graph denotes a potential
allocation of resources between peers $u$ and $v$.  The graph $G$ is
not necessarily complete although it is often assumed so in prior
work, but rather corresponds to a pre-computed topology. The overlay
model represents a snapshot of the system at a given time. The model
could apply to dynamic overlays by encompassing all logical
relationships during a time interval and then by weighting these edges
accordingly. 
An edge $\{u,v\}$ in $E$ can support the process of allocating
resources in both directions, i.e., $u$ can allocate resources to $v$
and $v$ can allocate resources to $u$. Therefore, every undirected
edge $\{u,v\}$ should be transformed into two directed edges $\toolarc
u v$ and $\toolarc v u$. The set of directed edges derived from $E$ is
denoted $E^\ast$.

The amount of rival resources that a peer $u \in V$ is able to offer
to other peers is termed $c(u)$, which does not exclusively mean
$c(u)$ \emph{different} data. The amount of resources that are given
by a peer $u$ to a neighbor $v$ corresponds to the weight $w(e)$
associated with the edge $e=\toolarc u v \in E^\ast$. For example the
peer $u$ reserves $w(e)$ bits per second to deliver video data to
$v$. The resource allocation can be represented by a weight function
$w: E^\ast \rightarrow \mathbb N$. Finally, each peer is also
associated with a demand, denoted $d(u)$, representing the amount of
resources that $u$ expects to receive from other nodes. In particular,
$d(u)$ is the minimal amount of resources that should be supplied to
$u$ in order to satisfy its quality of service requirements. While it
is trivial to add constraints on the links by associating a maximal
amount of resources that can be allocated from one peer to another, we
do not consider edge capacities in this paper. The only constraint for
the allocation $w(e)$ on the edge $e=\toolarc u v$ is either $c(u)$,
the amount of resources offered by $u$, or $d(v)$, the amount of
resources $v$ should receive.

\subsection{Resource allocation problems}

We study in this paper two instances of the problem of resource
allocation on the graph $G$. The first problem corresponds to the
stationary mode of the system, where nodes always have data to
contribute to their neighbours. The nodes can always satisfy the
resource allocation they have committed to. The problem can be
formulated as follows. \\

{\bf{Problem SRA}} (Stationary Regime Resource Allocation) Given an
overlay $G=(V, E^\ast)$ and capacity and demand distribution functions
$c(u)$ and $d(u)$, $u \in V$, determine the weight function $w: E^\ast
\rightarrow \mathbb N$ such that the demand $d(u)$ of all the nodes
$u$ can be satisfied.\\

We then relax the assumption on the availability of the non-rival
resources. We refer to the problem of resource allocation as the
$K$-\emph{Data-Capacitated Distribution Arborescence} (DCDA). We first
introduce the $1$-DCDA before generalizing to the $K$-DCDA. In the
$1$-DCDA, we consider that the resources that a node can contribute to
the system is contingent to data availability. The data can here be
seen as a file, a chunk or a stream. In particular, a node can
participate to the distribution in the overlay only if its demand has
been satisfied first.  The $1$-DCDA can be formally expressed as
follows. Given an overlay $G=(V, E^\ast)$, a source $s$ and a capacity
distribution function $c(u)$, $u \in V$, find the weight function $w:
E^\ast \rightarrow \mathbb \{0,1\}$ that maximizes the number of nodes
having a non-null incoming edge. The \emph{arborescence} rooted on $s$
formed by non-null weighted edges respects that, for all nodes $u$ in
the arborescence, the number of children of $u$ is not more than
$c(u)$.

We now generalize the problem to the case where the data are organized
into $K$ independent data units, \textit{e.g.}, $K$ chunks or $K$
different descriptions of a same video stream. The quality of service
$q(u)$ at a node $u$ is an increasing function of the number of data
units, therefore the demand $d(u)$ is $K$ and corresponds to a perfect
quality of service. The distribution of the data is organized into
separate trees $T_k, 0 \leq k \leq K$. For a node $u$ belonging to
$T_k$, its number of children in $T_k$ is noted $m_k(u)$. The problem
of the maximization of the overall quality of service can be written
as~:\\

{\bf{Problem $K$-DCDA}} Given an overlay $G=(V, E^\ast)$, a source $s$
and a capacity distribution function $c(u)$, $u \in V$, find the $K$
weight functions $w_k: E^\ast \rightarrow \{0,1\}, 0 < k \leq K,$ that
maximize the sum of quality of service $\sum_{u \in V} q(u)$. The
arborescences $T_k$ rooted on $s$ and formed by non-null weighted
edges in $w_k$ respect that, for all node $u$, $\sum_{k=1}^{K}
m_{k}(u) \leq c(u)$.\\

This problem specifies the demand as a boolean utility function on
each tree, which generally simplifies the problem of utility
maximization~\cite{chiang2007lod}. We do not try to maximize benefits
while spanning all nodes in the network, which is one of the most
studied problem in the literature. Rather, we aim to maximize the
number of fulfilled nodes. Finally, we note that the solution to the
SRA problem is the stationary regime solution of the $K$-DCDA problem
if the demand of all the nodes can be satisfied. In the next sections,
we show how to compute optimal and approximate solutions for these two
problems, and we analyze the performance of the resulting algorithms.

\section{Optimal Allocation in Stationary Regime}

Our goal here is to compute the allocation that maximizes the amount
of resources allocated between the peers in the overlay given their
demands and serving capacities. We will show that such an optimal
resource allocation can be computed in polynomial time. We will derive
our solution through a transformation mapping a problem related to
node-weighted graphs to a maximum flow problem on edge-weighted
graphs. Such a transformation is not
unusual~\cite{FrKiKo07,TaKlRa07,MaTwGk07}, however the problem tackled
here has never been formulated before with a graph-based model
featuring two weights for each vertex in the graph. The maximum flow
problem can then be solved with classic algorithms in polynomial
time. We emphasize that works dealing with similar problems have used
powerful but costly techniques to provide approximate
algorithms~\cite{streamingcapacity}. In comparison, our elegant
algorithm provides exact solutions in polynomial time.

\subsection{Transformation into a Flow Network}

We associate a \textit{network} $\mathcal N(G, c,d) =(V',E',w)$ to our
overlay $G$, featuring capacity and demand distribution functions
$c(u)$ and $d(u)$, $u \in V$. In particular, the set $V'$ contains a
sink $p$, a source $s$ and, for every peer $u \in V$, two vertices
$u^+$ and $u^-$. Let $V^+$ be the set $\{u^+: u\in V\}$ and $V^- =
\{u^-: u\in V\}$.  Formally, we have $V'=V^+ \cup V^- \cup \{s,p\}$.

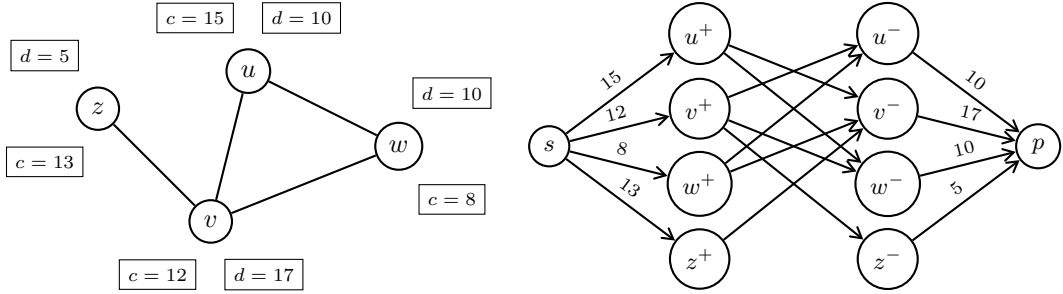
\begin{figure*}[!t]
  \centering
  \begin{tikzpicture}

\node[circle, thick, draw] (u) at (0,1) {$u$};
\node[circle, thick, draw] (v) at (-0.5,-1) {$v$};
\node[circle, thick, draw] (w) at (2,0) {$w$};
\node[circle, thick, draw] (z) at (-2,0.5) {$z$};

\node[draw, above right of= u] {\scriptsize{$d=10$}};
\node[draw, above left of= u] {\scriptsize{$c=15$}}; 
\node[draw, below right of= v] {\scriptsize{$d=17$}};
\node[draw, below left of= v] {\scriptsize{$c=12$}};
\node[draw, above right of= w] {\scriptsize{$d=10$}};
\node[draw, below right of= w] {\scriptsize{$c=8$}};
\node[draw, above left of= z] {\scriptsize{$d=5$}};
\node[draw, below left of= z] {\scriptsize{$c=13$}};

\draw[thick] (u) -- (v);
\draw[thick] (u) -- (w);
\draw[thick] (v) -- (z);
\draw[thick] (v) -- (w);





\node[circle, thick, draw] (source) at (4.,0) {\small{$s$}};

\node[circle, thick, draw] (vplus) at (6,0.5) {\small{$v^{+}$}};
\node[circle, thick, draw] (uplus) at (6,1.5) {\small{$u^{+}$}};
\node[circle, thick, draw] (wplus) at (6,-0.5) {\small{$w^{+}$}};
\node[circle, thick, draw] (zplus) at (6,-1.5) {\small{$z^{+}$}};

\node[circle, thick, draw] (vmoins) at (8.5,0.5) {\small{$v^{-}$}};
\node[circle, thick, draw] (umoins) at (8.5,1.5) {\small{$u^{-}$}};
\node[circle, thick, draw] (wmoins) at (8.5,-0.5) {\small{$w^{-}$}};
\node[circle, thick, draw] (zmoins) at (8.5,-1.5) {\small{$z^{-}$}};

\node[circle, thick, draw] (puits) at (10.5,0) {\small{$p$}};


\draw[thick,->, >= angle 60] (uplus) -- node[sloped,below,near start] {} (vmoins); 
\draw[thick,->, >= angle 60] (vplus) -- node[sloped,below,near start] {}(umoins);

\draw[thick,->, >= angle 60] (uplus) -- node[sloped,below,near start] {}(wmoins);
\draw[thick,->, >= angle 60] (wplus) -- node[sloped,below,near start] {}(umoins);

\draw[thick,->, >= angle 60] (vplus) -- node[sloped,below,near start] {}(wmoins);
\draw[thick,->, >= angle 60] (wplus) -- node[sloped,below,near start] {}(vmoins);

\draw[thick,->, >= angle 60] (zplus) -- node[sloped,below,near start] {}(vmoins);
\draw[thick,->, >= angle 60] (vplus) -- node[sloped,below,near start] {}(zmoins);

\draw[thick,->, >= angle 60] (source)-- node[sloped,above] {\scriptsize$15$} (uplus);
\draw[thick,->, >= angle 60] (source)-- node[sloped,above] {\scriptsize$12$} (vplus);
\draw[thick,->, >= angle 60] (source)-- node[sloped,above] {\scriptsize$8$} (wplus);
\draw[thick,->, >= angle 60] (source)-- node[sloped,above] {\scriptsize$13$} (zplus);

\draw[thick,->, >= angle 60] (umoins) -- node[sloped,above] {\scriptsize$10$} (puits);
\draw[thick,->, >= angle 60] (vmoins) -- node[sloped,above] {\scriptsize$17$} (puits);
\draw[thick,->, >= angle 60] (wmoins) -- node[sloped,above] {\scriptsize$10$} (puits);
\draw[thick,->, >= angle 60] (zmoins) -- node[sloped,above] {\scriptsize$5$} (puits);

\end{tikzpicture}
  \caption{Network transformation of an overlay containing four
    peers. The maximal capacities of the edges are also indicated.}
\label{fig:example_transformation}
\end{figure*}

The set of directed edges $E'$ includes three distinct subsets. The
first one contains $n$ edges from the source to each vertex in $V^+$,
where $n$ is the size of the vertex set $V$. The capacity of an edge
$\toolarc s {u^+}$ is the amount of resources $c(u)$ the peer $u$ can
supply. The second subset comprises $n$ edges from each vertex in
$V^-$ to the sink. Here, the capacity of an edge $\toolarc {u^-} p$
equals the demand $d(u)$. Finally, in the third subset of edges, we
assign one edge from $u^+$ to $v^-$ if there is an edge $(u
\rightarrow v) \in E^\ast$ in the original overlay graph. The capacity
of this edge is infinite\footnote{Adding a fixed link capacity here
  would be straightforward.}. Thus we can define $E'$ as
$E'=\{\toolarc s {u^{+}}, \toolarc {u^{-}} p : u \in V \}\cup
\{\toolarc {u^{+}} {v^{-}} : (u \rightarrow v) \in E^\ast\}$.  An
illustration of the transformation described here for the case of a
four peer overlay is shown in Figure~\ref{fig:example_transformation}.

Finally, let $f$ be a flow in $\mathcal N(G,c,d)$. A weight function
$w$ can be defined as: for every arc $\toolarc u v \in E^\ast$, set
$w\toolarc u v$ to $f \toolarc {u^+} {v^-}$.  The total amount of
allocated resources over $w$ is exactly the value of $f$ with respect
to both demand and capacity. The SRA resource allocation problem
becomes equivalent to a maximum flow problem on a bipartite graph.

\subsection{Optimal Resource Allocation}

In a maximum flow problem, the goal is to find the maximum value that
a flow between a single source and a single sink can achieve in a
network where each edge $e$ has a nominal capacity $c(e)$. Two famous
algorithms for computing the optimal solution in such instances are
Ford-Fulkerson and Edmonds-Karp. These algorithms have a time
complexity in $\mathcal O(|E|\cdot f)$ and $\mathcal O(n \cdot |E|^2)$,
respectively, where $n$ is the number of vertices of the flow network,
and $f$ the value of the maximum flow.


If the capacities exceed the demands, the value of a max flow is equal
to the sum of the demands because the capacity of the links from the
nodes in $V^+$ to the nodes in $V^-$ is infinite. Therefore, by
definition of flow conservation, if the value of $f$ is equal to the
sum of the demands, we obtain that $w$ reaches the maximum
demand. More generally, any maximum flow $f$ on $\mathcal N(G, c,d)$
allows to determine an associated weight function $w$ for $G$ such
that the demand for every peer is fulfilled if and only if the value
of $f$ is the sum of the demands. In other words, an answer to the
decision problem can be immediately deduced from a computation of the
maximum flow.

The max-flow problem can also be solved in a distributed way. This is
very interesting in practice since the nodes generally do not have a
global knowledge about the topology. Known distributed algorithms for
the max-flow problem in such a setting are based either on the
Ford-Fulkerson method~\cite{Barbosa96} or on the preflow-push
method~\cite{GoGuPe97}.  A basic implementation of such an algorithm
would allow the computation of an optimum resource allocation in any
peer-to-peer system.


\subsection{Discussion}

Bounded degree max flow problems have been shown to be
NP-complete~\cite{leblet08}, therefore our algorithm for computing the
optimal resource allocation can not be applied if an additional
constraint to the problem is to bound the number of neighbors to which
any peer can allocate resources. Yet, such a constraint is frequently
encountered in peer-to-peer systems, as discussed earlier. Hence,
another open problem in P2P capacity allocation consists of designing
an algorithm that would both maximize the resource allocation and
limit the degree of the resulting subgraph comprising only the
non-zero weighted edges.

Through the algorithm described thus far one can determine if there is
an optimal allocation that fulfills the demands of all nodes in the
overlay. However, if all nodes can not be fulfilled, this algorithm
cannot compute the allocation of resources that maximizes the number
of fulfilled nodes in the overlay. The algorithm presented in the next
section is able to find an allocation that maximizes the number of
fulfilled nodes.

\section{Data-Capacitated Distribution Problem}

We now include non-rival resources in the resource allocation problem,
and we compute the capacity of the system under data availability
constraints. A peer can not allocate any of its uplink bandwidth if it
does not have data to transmit first. The non-rival resources are a
set $K$ comprising independent data units. Data are roughly equivalent
in size. The quality of service associated with a peer is then a
function of the number of received data. We denote by $q(u)$ the
service quality for a node $u$. The quality of service is a increasing
function of the number of data units received by the peer.

Each data $k \in K$ is served to nodes on a separate arborescence
$T_k=(V_k,E_k)$, a directed tree rooted at $s$ where $V_k \subseteq V$
and $E_k \subseteq E^\ast$. The children of a client $u \in V_k$ are
denoted by $N_k(u)$, the number of children by $m_k(u)$. The multiple
tree construction takes into account the aforementioned constraint on
upload capacity of node $u$, i.e., $\sum_{k\in K} m_k(u) \leq c(u),
\forall u \in V$.

As stated in the problem DCDA, we are interested in maximizing the
overall quality of service in the overlay. Here, we define this
quantity to be the sum of the qualities of service $q(u)$ experienced
by all clients. Our model can support alternative definitions of the
overall quality of service, as ensuring fairness among the clients or
maximizing the number of clients up to a given quality threshold. We
show below that the $K$-DCDA is NP-complete, even for $K = 1$.

\subsection{NP-Completeness of $K$-DCDA}

A formal formulation of the decision problem related with $k-$DCDA is: \\
{\sc Instance :} A graph $G=(V,E^\ast)$ with $V$ the set of vertices and $E$ the set of edges, a root $s\in V$, a positive integer $K$, a capacity function $c:V \longrightarrow {\mathbb N}$ and a positive integer $\Gamma$.\\
{\sc Question :} Do there exist $K$ arborescences $\left
  (T_k=(V_k,E_k) \right )_{1\leq k\leq K}$ rooted in $s$ such that:
\begin{itemize}
\item[(1)] for any $k$, we have $V_k\subseteq V$ and if $(u \rightarrow
  v)$ is an edge from $T_k$ then $(u \rightarrow v)$ is an edge of $G$,
\item[(2)] for any vertex $u\in V$, the sum of its outdegrees is
  lower or equal to its capacity, i.e., $\sum_{k=1}^K m^+_{T_k}(u)
  \leq c(u)$,
\item[(3)] the total number of vertices belonging to the
  arborescences is greater or equal to $\Gamma$, i.e., $\sum_{k=1}^K
  |V_k| \geq \Gamma$.
\end{itemize}



We now provide a proof of the NP-completeness of $K$-DCDA using a
reduction to the famous 3-SAT problem.

{\noindent\bf 3-SAT\\} {\sc Instance :}
Set $U$ of variables and a collection $C$ of clauses over $U$ such that each clause $c\in C$ has $|c|=3$.\\
{\sc Question :} Is there a truthful assignment for $C$?

\begin{theorem}
  {\bf $K$-DCDA} is NP-complete even for $K=1$.
\end{theorem}

\bpr Given an instance of $K$-DCDA Problem and a family $\left
  (T_k=(V_k,E_k) \right )_{1\leq k\leq K}$ of $K$ arborescence rooted
in $s$, verifying that this family is a valid one is clearly
polynomial in the size of the problem: hence the $K$-DCDA problem
belongs to NP.


Now, given an instance of the 3 SAT problem comprising
$U=\{x_1,\cdots,x_n\}$ a set of variables and
$C=\{C_1,\cdots,C_{|E|}\}$ a set of clauses on $U$ where
$C_j=x_j^1\vee x_j^2\vee x_j^3 $, we define an instance of the
$K$-DCDA problem as follows.  Recall that for any $1\leq j \leq |E| $
and any $1\leq l\leq 3$, we have that there exists $1\leq i\leq n$
such that $x_j^l\in \{x_i, \overline{x_i}\}$. Let $V=\{s\}\cup \{i,
x_i, \overline{x_i} : 1\leq i \leq n\}\cup \{C_1,\cdots,C_{|E|}\}$ and
let $E'=\{\{s,i\} : 1\leq i \leq n\}\cup \{\{i,x_i\},
\{i,\overline{x_i}\} : 1\leq i \leq n\} \cup \{ \{x_j^l,C_j\}: 1\leq j
\leq |E|, 1\leq l\leq 3\}$.  For $1\leq i \leq n$ and for $1\leq j\leq
|E|$, the capacity function is defined as $c(s)=n$, $c(i)=1$,
$c(x_i)=c(\overline{x_i})=|E|$ and $c(C_j)=0$.  Finally, we define
$\Gamma$ as $1+2n+|E|$.  Clearly the instance of the $K$-DCDA problem
can be constructed in polynomial time in the size of the 3-SAT
instance.  We claim that there exists an arborescence $T=(V',F)$
solving our instance of the problem $K$-DCDA if and only if there
exists a truthful assignment for $C$.

For the forward implication, assume that there exists an arborescence
$T=(V',F)$ fulfilling conditions~$(1)$ to~$(3)$ of the problem
$K$-DCDA.  As $K=1+2n+m=|V'|$ and as for any $1\leq k \leq n$, we have
$c(k)=1$ and $c(j)=0$ for $1\leq j \leq |E|$, it follows that
$|\{x_i,\overline{x_i}\}\cap V'|=1$ for any $1\leq i \leq n$.  We
define the assignment function $\varphi$ as follows : $\varphi(x_i)$
is set to True if $x_i\in V'$ and False if $\overline{x_i}\in V'$.
But now, as $|V'|=1+2n+|E|$ and as for any $1\leq i \leq n$ it holds
$|\{x_i,\overline{x_i}\}\cap V'|=1$, we obtain that for any $1\leq
j\leq |E|$, $C_j \in V'$ and thus that there exists a vertex $x'_j$ in
$\{x_i,\overline{x_i} : 1\leq i \leq n\}$ such that $(x'_j,C_j)$ is an
edge of $T$.  But now, by definition of $\varphi$, we obtain that the
literal associated to $x'_j$ has a True value and thus we obtain that
the clause $C_j$ has also a true value, and thus that $\varphi$ is a
truth assignment for $C$.

For the backward implication, assume that we have a truth assignment
$\varphi$ for $C$.  We define $U'$ the set of true litterals for
$\varphi$, that is $U'=\{x_i : x_i \in U, \varphi(x_i)=True\}\cup
\{\overline{x_i} : x_i \in U, \varphi(x_i)=False\}$.  Now let
$V'=\{s\}\cup \{1,\cdots,n\} \cup U' \cup C$, clearly we have
$|V'|=1+2n+|E|$.  As $C$ is True, this means that for any $1\leq j\leq
|E|$, there exists at least one literal $y_j \in C_j$ such that
$\varphi(x_j)=True$.  We denote by $y_j$ one literal from $C_j$ which
is True by $\varphi$.  We define $F=\{(s,i): 1\leq i \leq n\}\cup \{
(i,x_i) : 1\leq i\leq n, x_i \in U'\}\cup \{ (i,\overline{x_i}) :
1\leq i\leq n, \overline{x_i} \in U'\}\cup \{(y_j,C_j) : 1\leq j \leq
|E|\}$.  As by definition of $y_j$, the literal $y_j$ is set to True
and by the definition of $F$, it is obvious that $D=(V',F)$ is an
arborescence rooted in $s$ and that edges from $D$ are also edges from
$G$.  Now we remain with the capacity constraint.  Clearly we have
$m_D(s)=n$, for any $1\leq j \leq |E|$, $m_D(C_j)=0$.  Now, as for any
$1\leq i \leq n$, we have $|\{x_i,\overline{x_i}\}\cap U'|=1$, we
obtain that $m_D(k)=1$.  Moreover, for any $1\leq i \leq n$, we have
both $m_D(x_i)\leq |E|$ and $m_D(\overline{x_i})\leq |E|$, thus $D$ is
an arborescence fullfilling conditions~$(1)$ to $(3)$ and having
$\Gamma$ elements.  \epr

Note that the backward implication above only considers the case $K =
1$ since showing that 1-DCDA is NP-complete also implies that $k-DCDA$
is NP-complete.



\subsection{$1$-DCDA Problem Decomposition}

As the $K$-DCDA problem is NP-complete even for $K=1$, we focus now on
the particular instance of the $1$-DCDA problem where the quality of
service is a binary function. The peers either fulfills their demand
$d(u)= 1, \ \forall u \in V$, or not. We propose a decomposition of
the $1$-DCDA problem into a master problem and several subproblems,
which can be solved efficiently in polynomial time. We introduce first
the concept of \emph{level}. The vertex $s$ corresponds to the only
vertex at level $0$; the vertices adjacent to $s$ are at level $1$,
the vertices adjacent to those at level 1 are at level $2$, and so
forth.  The level of a vertex therefore represents its distance (in
terms of hops) to vertex $s$ in the tree. We denote by $J =
\{1,2,3,\cdots, n-1\}$ the set of possible levels. We also denote by
$V_s$ the set of nodes $V \setminus \{s\}$.

Let $x \in \{0,1\}^{(n-1)^2}$ be a matrix defined as:
\begin{equation*}
x^j_v = \left\{
\begin{array}{ll}
1 & \textrm{ if } v \textrm{ is at level } j,\\
0 & \textrm{ otherwise,}
\end{array}
\right.
\end{equation*}
for all $v \in V_s$ and $j \in J$.  Furthermore, let $y \in
\{0,1\}^{|E|(n-1)}$ be the matrix defined as:
\begin{equation*}
y^j_e = \left\{
\begin{array}{ll}
1 & \textrm{ if } e \textrm{ is selected from level } j-1 \textrm{ to level } j,\\
0 & \textrm{ otherwise,}
\end{array}
\right.
\end{equation*}
for all $e \in E$ and $j \in J$.  Then, the $1$-DCDA problem is equivalent to
the following mixed-integer linear program P1
\[\text{P1} : \, \, \max z(x) = \sum\limits_{j=1}^{n-1} \sum\limits_{v \in V_s} x_v^j \, , \quad \text{s.t.} \]
\vspace{-0.25cm}
\begin{align}
\label{inq:vertex:cover} & \sum\limits_{j=1}^{n-1} x_v^j \leq 1 \, , \, \textrm{for } v \in V_s,\\
\label{inq:degree:s} & \sum\limits_{v \in V_s} x^1_v \leq c(s),\\
\label{inq:degree:t:t+1} & \sum\limits_{v \in V_s} x_v^j - \sum\limits_{v \in V_s} c(v)x_v^{j-1} \leq 0 \, , \, \textrm{for } j \in J\setminus\{1\},\\
\label{inq:edge:assign} & \sum\limits_{j=1}^{n-1} y_e^j \leq 1 \, , \,  \textrm{for } e \in E,\\
\label{inq:edge:left:s} & \sum\limits_{e \in \delta(s)} y_e^1 - c(s) \leq 0,\\
\label{inq:edge:left} & \sum\limits_{e \in \delta(v)} y_e^j - c(v)x_v^{j-1} \leq 0 \, , \, \textrm{for } v \in V_s,\ j \in J\setminus\{1\},\\
\label{inq:edge:right} & \sum\limits_{e \in \delta(v)} y_e^j - x_v^j = 0 \, , \, \textrm{for } v \in V_s,\ j \in J, \\
\label{inq:vertex:integral} & x \in \{0,1\}^{(n-1)^2},\\
\label{inq:edge:integral} & y \in \{0,1\}^{|E|(n-1)}.
\end{align}

The assignment of vertex $v \in V_s$ to at most one level is expressed
by inequalities \eqref{inq:vertex:cover}. Inequalities
\eqref{inq:degree:s} and \eqref{inq:degree:t:t+1} bound from above the
number of vertices at level $j+1$, based on the number of vertices at
level $j$ and the node capacity function $c$.  Inequalities
\eqref{inq:edge:assign} guarantee that edge $e \in E$ is selected at
most once in the induced tree.  Inequalities \eqref{inq:edge:left:s}
and \eqref{inq:edge:left} ensure that vertex $v \in V$ at level $j$ is
adjacent to at most $c(v)$ vertices at level $j+1$, whereas
inequalities \eqref{inq:edge:right} ensure that vertex $v \in V_s$ at
level $j$ is adjacent to exactly one vertex at level $j-1$.

This model can be reformulated without the $y$ variables using
Benders' decomposition~\cite{benders1962pps}.  The main principle of
this decomposition consists of separating the variables of the
problem. A master problem, still NP-complete, is in charge of
determining a solution for one variable, while the sub-problems are
responsible to complete the assignment on the other variables. If this
assignment is possible, the whole problem is solved, otherwise a new
constraint is added to the master problem, which makes its computation
quicker.

Now, let $X = \left\{ x \in \mathbb{R}^{(n-1)^2} : \, x \textrm{
    satisfies } \eqref{inq:vertex:cover}-\eqref{inq:degree:t:t+1}
  \textrm{ and } \eqref{inq:vertex:integral}\right\}$.  
Moreover, let (\ref{inq:edge:left}.$j$) and (\ref{inq:edge:right}.$j$) denote
inequalities \eqref{inq:edge:left} and \eqref{inq:edge:right} for a
specific value $j$ in $J\setminus\{1\}$ and $J$, respectively. Then,
let
\[Y(1) = \{y^1 \in \mathbb{R}^{|E|} : y^1 \textrm{ satisfies }
  \eqref{inq:edge:left:s}, (\ref{inq:edge:right}.1) \textrm{ and } y^1
  \in \{0,1\}^m\},\] \noindent while for $j\in
J\setminus\{1\}$ let
\[Y(j) = \{y^j \in \mathbb{R}^{|E|} : y^j \textrm{ satisfies } (\ref{inq:edge:left}.j), (\ref{inq:edge:right}.j) \textrm{ and } y^j \in \{0,1\}^m\}.\]


Finally, the program P1 can be rewritten as
\begin{equation}
  \max\limits_{x \in X} z(x) + \zeta(s,x^1) + \sum\limits_{j=2}^{n-1} \zeta(x^{j-1},x^j) \, ,
  \label{eqn:master_slave_program}
\end{equation}
where the subproblems have no incidence on the value of the final
solution, therefore they can be abusively written as:
\begin{eqnarray}\label{eq:subpb:s}
  \zeta(s,x^1) & = & \max_{y^1 \in Y(1)} 0 \, , \\
\label{eq:subpb:t}
  \zeta(x^{j-1},x^j) & = & \max_{y^j \in Y(j)} 0 \, , \, \text{for } \, j\in J\setminus\{1\} \, .
\end{eqnarray}

The idea behind the decomposition in (\ref{eqn:master_slave_program})
is that a master problem generates a solution where the nodes are
assigned to levels, and then the individual sub-problems verify if it
is indeed possible to find edges linking the nodes at a given level
with the nodes at the next level while respecting the node capacity
function $c$. Next, we show that these sub-problems can be solved in
polynomial-time.

Consider an undirected graph $G^j = (V^j,E^j)$, a partition
$\{L^j,R^j\}$ of $V^j$ and a function $b : L^j \longrightarrow
\mathbb{N}$.  A {\em semi-perfect $b$-matching of $G^j$} is a subset
$M$ of edges of $G^j$ such that every vertex $v$ in $L^j$ is incident
with at most $b_v$ edges of $M$ and every vertex in $R^j$ is incident
with exactly one edge of $M$. In our case, the number of used links
from nodes in $L^j$ should not be higher than the capacity of this
node, while only one link should be used to reach the nodes in $R^j$.
Let $M$ be a semi-perfect $b$-matching of $G^j$. Its {\em incidence
  vector} $\chi$ is the $\{0,1\}$-vector in $\mathbb{R}^{E^j}$
satisfying
\begin{equation*}
\chi^M_e = \left\{
\begin{array}{ll}
1 & \textrm{ if } e \in M^J,\\
0 & \textrm{ if } e \in E \setminus M^J.
\end{array}
\right.
\end{equation*}
The incidence vectors of semi-perfect $b$-matchings of $G^J$ are solutions
to the following system of linear inequalities
\begin{align}
\label{eq:sp:matching:left} & x(\delta(v)) \leq b_v && \textrm{ for } v \in L^j,\\
\label{eq:sp:matching:right} & x(\delta(v)) = 1 && \textrm{ for } v \in R^j,\\
\label{eq:sp:matching:non-negativity} & x_e \geq 0 && \textrm{ for } e \in E^j.
\end{align}

A polyhedron $P$ is {\em integral} if $P$ is the convex hull of the
integral vectors in $P$. A pointed polyhedron $P$ (\textit{i.e.},
containing at least one extreme point) is integral if and only if each
vertex is integral~\cite{schrijver2003cop}.  In the next lemma, we
show that the polyhedron defined by inequalities
\eqref{eq:sp:matching:left}-\eqref{eq:sp:matching:non-negativity} is
integral.

\begin{lemma} \label{lemma:semi-perfect:matching polytope}
The polyhedron
\begin{equation*}
SPMP(G^j,b) = \{x \in \mathbb{R}^{E^J} : x \textrm{ satisfies } \eqref{eq:sp:matching:left}-\eqref{eq:sp:matching:non-negativity}\}
\end{equation*}
is integral, providing it is not empty and $G^j$ is bipartite.
\end{lemma}
\bopr Assume $G^j$ is bipartite and $SPMP(G^j,b) \not= \emptyset$.
Let $H^j$ be the incidence matrix of $G^j$ which is known for being
totally unimodular~\cite{motzkin}. Matrix $H^j$ can be partitioned
into $H^{L^j}$ and $H^{R^j}$, where $H^{L^j}$ and $H^{R^j}$ are
composed of the rows of $H^j$ indexed by the vertices of $L^j$ and
$R^j$, respectively.  If $x^\angle$ denotes the vector of slack variables of
\eqref{eq:sp:matching:left}, then system
\eqref{eq:sp:matching:left}-\eqref{eq:sp:matching:non-negativity} can
be rewritten as
\begin{equation*}
A'x' = \begin{pmatrix} H^{L^j} & I_{|L^j|}\\ H^{R^j} & 0 \end{pmatrix} \begin{pmatrix} x\\ x^\angle \end{pmatrix} = \begin{pmatrix} b\\ 1_{|R^j|}\end{pmatrix} = b',~ x'\geq 0
\end{equation*}
From $H^j$ being totally unimodular, we easily conclude that so is
matrix $A'$.  Since $b'$ is an integral vector, the polyhedron
$SPMP(G^j,b)$ is therefore integral~\cite{hoffmankruskal}.  \epr


It is straightforward to see that each of the subproblems
\eqref{eq:subpb:s}-\eqref{eq:subpb:t} corresponds to determining
whether a semi-perfect $b$-matching exists on a graph induced by the
vertices between two consecutive levels. In fact, consider any $j \in
J$, and define $L^j = \{v \in V : x_v^{j-1} = 1\}$ and $R^j = \{v \in
V : x_v^j = 1\}$. (If $j=1$, then $L^1$ is reduced to vertex $s$.)
The subgraph $G^j$ of $G$ clearly is bipartite because of inequalities
\eqref{inq:vertex:cover}.

Using Lemma \ref{lemma:semi-perfect:matching polytope} and Farkas'
Lemma~\cite{farkas1902} (or duality in linear programming), each of
the subproblems \eqref{eq:subpb:s}-\eqref{eq:subpb:t} has a feasible
solution if and only if
\begin{equation} \label{eq:extreme:ray}
 u^J(Cx^{j-1} + x^j) \geq 0
  \quad \textrm{ for every extreme ray } u \textrm{ of } C(j),
\end{equation}
where $C(j) = \{(u^{j-1} u^j) \in \mathbb{R}^{n_l+n_r} : (u^{j-1}
u^j)^TH^j \geq 0,\ u^{j-1} \geq 0\}$ and $H^j$ is the incidence matrix
of the subgraph $G^j$.  Therefore, the integer linear programming
formulation
\begin{equation*}
  \max\limits_{x,y} z(x) \textrm{ s.t. } \eqref{inq:vertex:cover}-\eqref{inq:edge:integral}
\end{equation*}
is equivalent to solving
\begin{equation*}
  \max\limits_{x} z(x) \textrm{ s.t. } \eqref{inq:vertex:cover}-\eqref{inq:degree:t:t+1},\eqref{inq:vertex:integral},\eqref{eq:extreme:ray},
\end{equation*}
with the separation problem of inequalities \eqref{eq:extreme:ray}
being solvable in polynomial time (it reduces to solving linear
programs).

\section{Heuristic Resource Allocation Algorithms}

The previously described decomposition aims to reduce the computation
time of the exact solution. Even if the decomposition is promising, it
still cannot solve the original $K$-DCDA problem. In addition, the
exponential nature of the problem makes that it is not reasonable to
expect results for large instances of the problem. Yet, peer-to-peer
architectures make sense when the number of clients is
large. Therefore we are looking for heuristics running in
polynomial-time and determining solutions that are not far from the
exact solution.

Several generic approaches have proved to be especially efficient in
searching solutions to NP-hard optimization problems. For example,
\emph{genetic algorithms} use techniques inspired by evolutionary
biology to compute an almost optimal solution from a set of valid
non-optimal instances~\cite{haupt2004practical}. The computation is
based on successive steps. At each step, a new generation of solutions
is produced from the previous generation. The main idea is that these
successive generations are expected to evolve toward better
solutions. Various optimization techniques have been studied to
improve the performance of genetic algorithms, but, as they are
inherently generic, genetic algorithms are commonly outperformed by
dedicated heuristics applying on a given problem. Nevertheless, we
have implemented a generic algorithm for the $K$-DCDA, which allows to
compare other heuristics, and to provide an overview of the solution
for a large instance of the problem.

We have also designed a heuristic algorithm described in
Algorithm~\ref{alg:greedy}.  For each non-rival resource $k$, a node
can be in one state among four: \emph{dead$_k$} if it is served in
$T_k$ but it has no more resource to allocate, \emph{fulfilled$_k$} if
the node is served in $T_k$ and it can serve still one of its
neighbors, \emph{accessible$_k$} if it is not served yet in $T_k$ but
one of its neighbors is, and \emph{not accessible$_k$} otherwise. At
each step, an arborescence $T_k$ and a node $u$ being in the
\emph{accessible$_k$} state are chosen. Then a node in
\emph{fulfilled$_k$} is selected to serve $u$ in $T_k$. The algorithm
ends when no node is \emph{accessible$_k$} in any arborescence $T_k$.

\begin{algorithm}

\caption{Greedy Algorithm}\label{alg:greedy}
\dontprintsemicolon
\SetKwInOut{Input}{Input}
\SetKwInOut{Output}{Output}
\Input{a graph $(V,E)$, a source $s\in V$, a capacity function $c:V\rightarrow \mathbb {N}$}
\Output{a set of $K$ arborescence $T_k=(W_k,E_k)$}

$W_k\leftarrow\{s\}$ ; $E_k\leftarrow \emptyset$\;
$Dead_k\leftarrow \emptyset$\;
$Fulfilled_k \leftarrow \{s\}$\;
$Access_k\leftarrow N(s)$\;
$Not\_Acc_k\leftarrow V\setminus (Access_k\cup Dead_k \cup  Fulfilled_k)$\;
\While{$\exists k$ s.t. $Access_k\neq\emptyset$}
{
let $T_k$ a random arborescence with $Access_k \neq \emptyset$\;
\ForEach{$node \in Access_k$}
	{
	$nb\_not\_acc \leftarrow |N_k(node)\cap Not\_Acc_k|$\;
	$score(node)\leftarrow\min(nb\_not\_acc,c(node))$\;
	}
let $sel\_node$ the node with max. score\;
$Poss\_parent\leftarrow N(sel\_node)\cap Fulfilled_k$\;
let $par$ the node in $Poss\_parent$ with max. capacity\;
add $sel\_node$ to $W_k$\;
add edge $par \rightarrow sel\_node$ to $E_k$\;
$c(par) \leftarrow c(par)-1$\;
\eIf {$c(sel\_node)> 0$}
{move $sel\_node$ to $Fulfilled_k$\;}
{move $sel\_node$ to $Dead_k$\;}
\If {$c(par)=0$}
{
move $par$ from $Fulfilled_{k'}$ to $Dead_{k'}, \forall k'$\;
}
update $Access_{k'}, \forall k'$\;
update $Not\_Acc_{k'},\forall k'$\;
}
\end{algorithm}

The choice of the arborescence and the node to serve is crucial. In
Algorithm~\ref{alg:greedy}, we describe the ``\emph{greedy}'' approach
that has given so far the best results during our simulations. In
line~7, we use a uniform random choice to pick an arborescence having
a non-null set of accessible nodes. This uniform random choice
guarantees no privileged non-rival resource. Once an arborescence
$T_k$ is determined, a node in \emph{accessible$_k$} should be
chosen. For every candidate node $u$, we evaluate the number of
neighbors $u$ is able to serve in $T_k$, that is, the \emph{score} of
$u$ depends on its available capacity and on the number of its
neighbors in \emph{not accessible$_k$}. This part of the algorithm is
described in lines~8~to~11. Finally, the algorithm determines a parent
for $u$. Our approach consists of selecting the node that has the
largest amount of available resource (lines~12~and~13). The remaining
of the algorithm deals with state updating (lines~17~to~24).

This algorithm ensures a greedy construction of every tree, and
supports efficient distributed implementations. In the next part, we
evaluate two variants: the ``\emph{random}'' algorithm where the node
to serve is chosen at random instead of using a score, and the
``\emph{pre-fixed}'' algorithm where every node $u$ assigns a fixed
capacity for every tree $c_k(u), \sum_{k\leq K} c_k(u)=c(u)$, then the
algorithm computes $K$ greedy trees.


\section{Performance Analysis}








The goal of this simulation is threefold: evaluating the influence of
non-rival resources on the capacity of peer-to-peer networks,
estimating the ratio of fulfilled nodes for representative overlays,
and examining heuristic performances.

\subsection{Configuration}

Many recent works, including in standards organization, have dealt
with matching overlay networks and Internet. These network-friendly
overlays are fairly representative of the next generation of \textit{a
  priori}-constructed overlays, yet the degradation of performances
resulting from this non-optimal construction is still unknown. These
overlays illustrate the interest of our work, so we use in our
simulations the proximity of peers into an underlying Internet to
build the overlays. The underlying network is a matrix of latencies
between $2,500$ nodes from the Meridian project\footnote{Measurements
  have been done in May 2004, more information on
  \texttt{http://www.cs.cornell.edu/People/egs/meridian/}}.  For each
run, we choose randomly $n$ nodes, then, for each node, we determine
its $\kappa$ closest nodes among the selected nodes, and we establish
a connection between them. Therefore, the minimal degree of a node is
$\kappa$. Note that a node can be among closest neighbors of more than
$\kappa$ nodes, so its degree can be larger than $\kappa$. As a
result, the overlay is a bi-directional $\kappa$-nearest neighbor
graph built from a realistic set of nodes in the Internet. To
eliminate random effects, more than $20$ different instances are
tested for each measure.

We measure the ratio of allocated resources. In our context, the
demand of peers is the same for all peers, \textit{i.e.,} every peer
would like to receive the same amount of resources. We set $K$ to $3$,
and we use $d(u)=3,\forall u \in V$ for peers' demand in the
stationary regime. Hence, it is possible to compare both resource
allocation problems, in stationary regime and when $K$ non-rival
resources should be delivered. The average capacity is fixed to
3. Note that the average capacity being equal to the average demand,
the system is thus pushed to its limit: a ratio of allocated resources
equal to 1 means a perfect allocation of resources with no capacity
loss.

We show the results obtained by four heuristic algorithms. The
\emph{GA} algorithm corresponds to an implementation of a Genetic
Algorithm, with an initial population of $150$ basic solutions, and
$300$ steps. The \emph{greedy} algorithm is described in
Algorithm~\ref{alg:greedy}. Finally, both \emph{random} and
\emph{pre-fixed} heuristic algorithms have been previously introduced.

\subsection{Large Instances}

Our first focus is on the differences between SRA and $K$-DCDA problem
solutions, and an overview of the ratio of fulfilled nodes in large
representative overlays.

\subsubsection{Population Size}

\begin{figure}
  \centering
  \includegraphics[width=0.8\columnwidth]{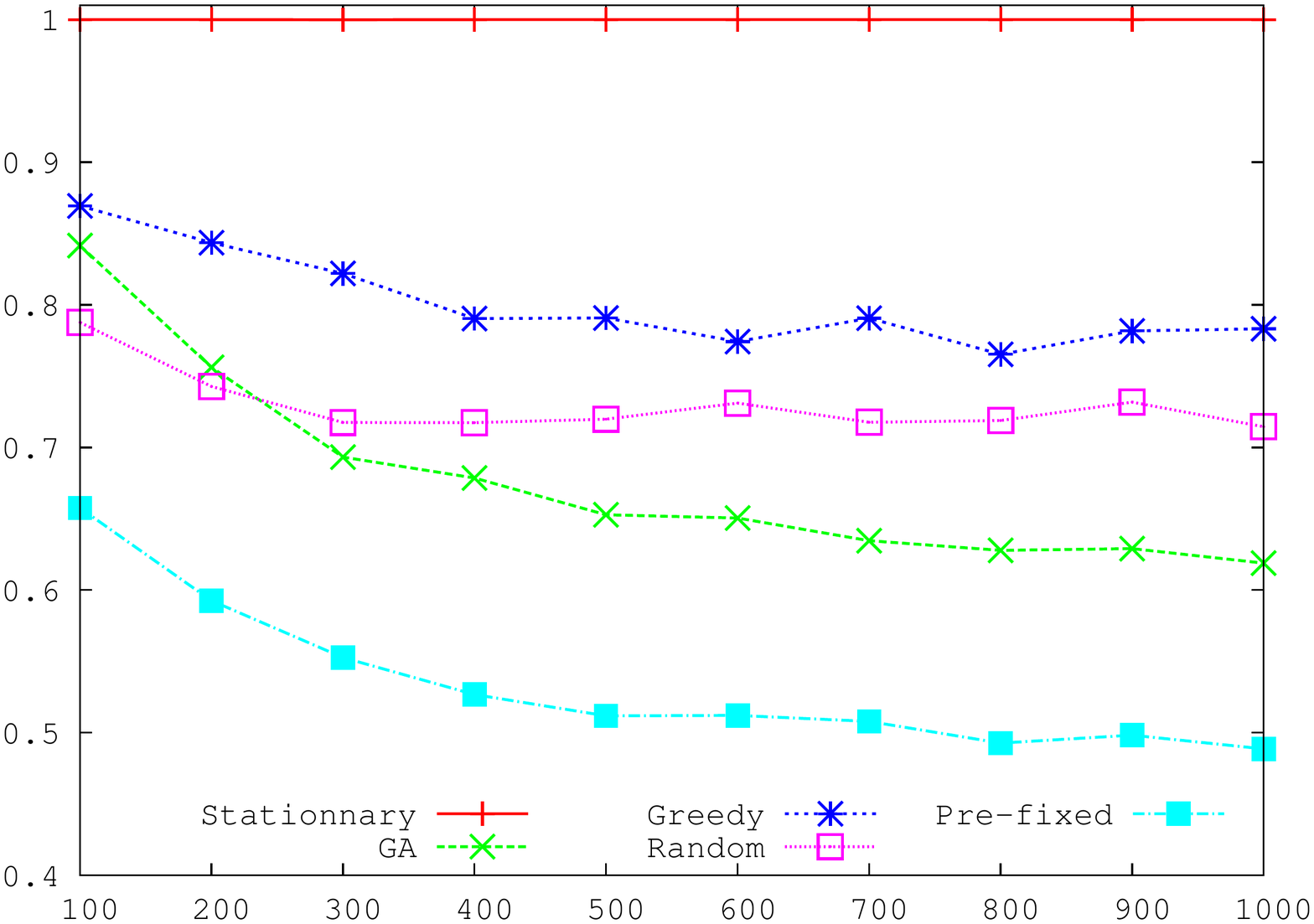}
  \caption{Ratio of allocated resources vs. population size $n$ for
    large instances}
  \label{fig:large}
\end{figure}

The number of peers $n$ varies from $100$ to $1,000$, the range of
capacities is from to $2$ to $4$, the parameter $\kappa$ is set to
$6$. Results are plotted in Figure~\ref{fig:large}.

In these configurations, there exists always a resource allocation
that fulfills all peers in the stationary regime. On the contrary, all
heuristic algorithms fail to find any perfect resource allocation with
non-rival resources. Although these algorithms do not guarantee any
optimal solution, we conjecture that non-rival resources add a
constraint that not only makes the best allocation harder to
determine, but also prevents some peers to fully use their capacities.

The performances of the \emph{GA} algorithm degrade quicker than other
heuristic algorithms. Intuitively, the wider is the solution space,
the worse are the performances of genetic algorithms. As can be
expected, \emph{GA} does not really perform better than efficient dedicated
heuristic algorithms.

A clear hierarchy is revealed among the three other algorithms. The
\emph{greedy} algorithm outperforms both other variants. We emphasize
the bad performances of the \emph{pre-fixed} algorithm, which fulfills
less than half of the peers when $n$ is $1,000$, while almost four
fifth of resources can be allocated by the \emph{greedy}
algorithm. This huge difference demonstrates that a not-so-clever
resource allocation can significantly degrade the performances of an
overlay.


 
\subsubsection{Network Density}

\begin{figure}
  \centering
  \includegraphics[width=0.8\columnwidth]{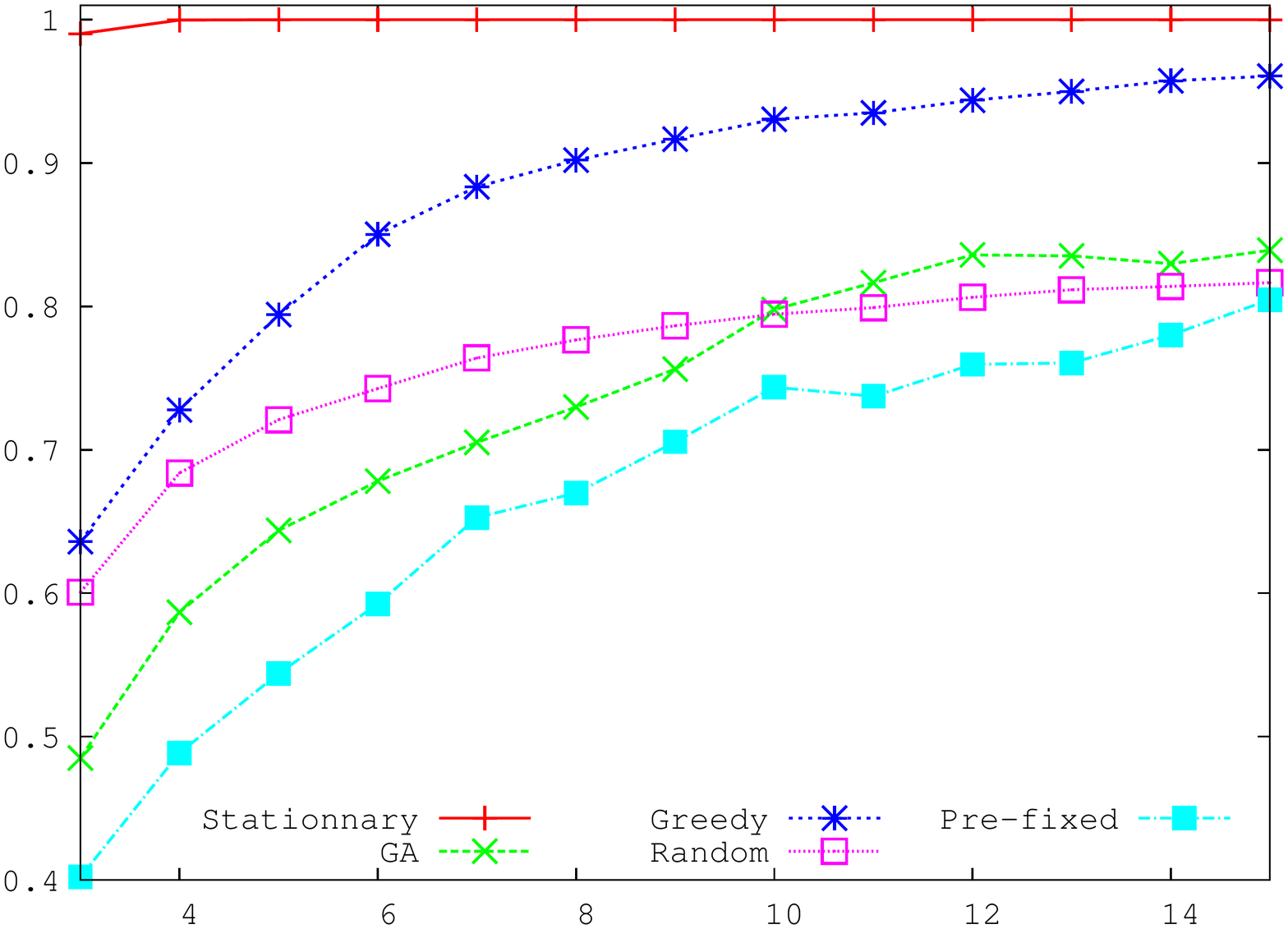}
  \caption{Ratio of allocated resources vs. minimum degree $\kappa$
    for large instances}
  \label{fig:deg}
\end{figure}

We consider now various overlay densities. The minimal degree $\kappa$
varies from $3$ to $15$, while the population size $n$ is fixed to
$200$. Results are in Figure~\ref{fig:deg}.

In stationary regime, the previous results are confirmed: even for
sparse overlays ($\kappa=4$), a valid resource allocation can fulfill
all nodes. This result highlights the importance of resource
allocation strategies, and the benefits one can expect from them on
any \emph{a priori}-constructed overlay.

With non-rival resources, this optimality can unfortunately not be
reached by our heuristics, though the performances are excellent for
dense networks. When $\kappa$ grows, the set of peers that are
candidate to be served enlarges, and the random choice becomes
naturally worse than a specific policy. Hence, the \emph{random}
strategie tends to underperform.

\subsection{Small Instances}

\begin{figure}
  \centering
  \includegraphics[width=0.8\columnwidth]{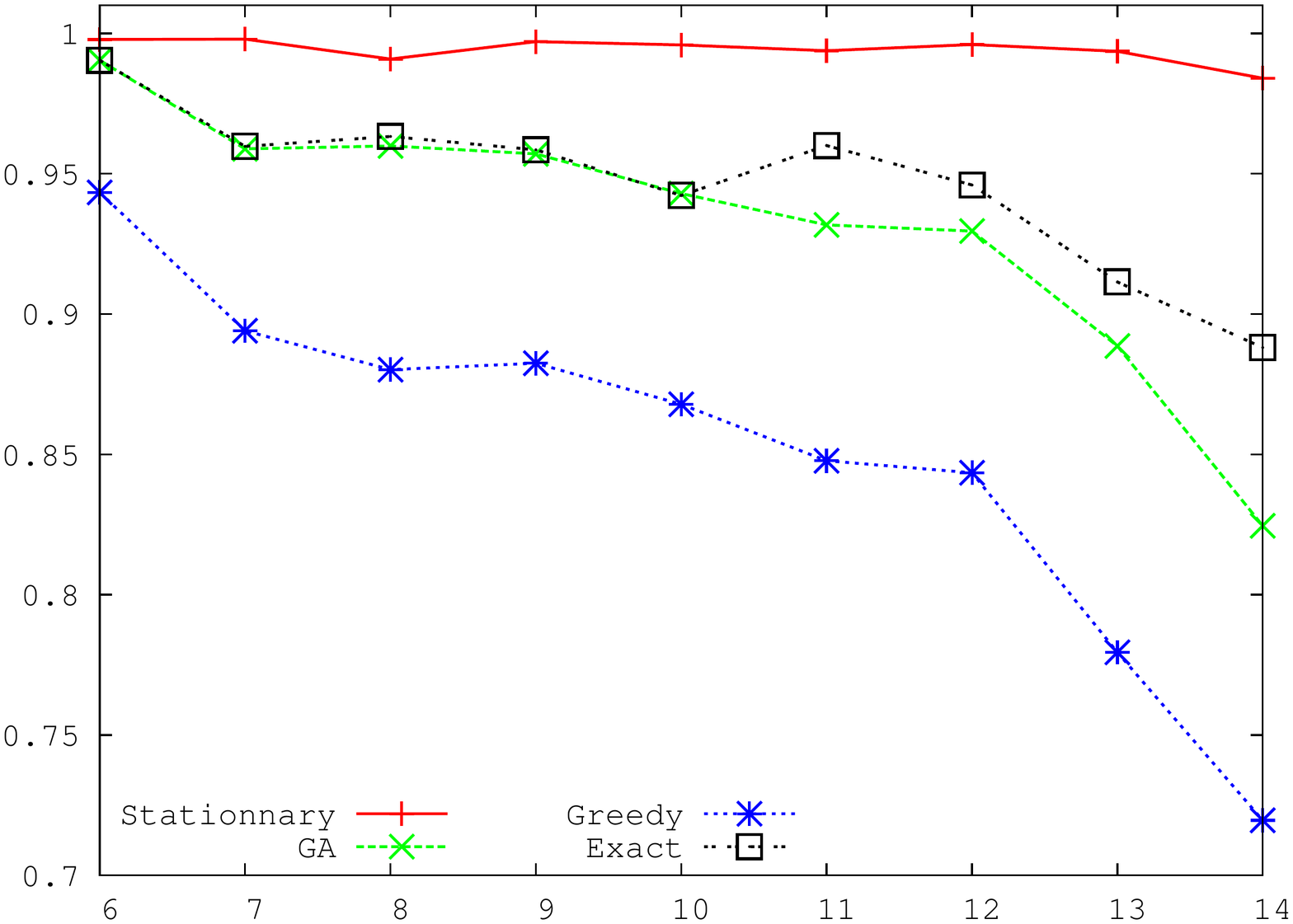}
  \caption{Ratio of allocated resources vs. population size $n$ for
    small instances}
  \label{fig:small}
\end{figure}

We now build small instances with $n$ from $6$ to $15$ nodes. In this
context, $\kappa$ is fixed to $3$ and the range of upload capacities
is from $0$ to $6$. On such small instances, exact solutions can be
computed in a reasonable time. Results in Figure~\ref{fig:small} aim
to provide a slight indication of the overall performances of our
heuristics. We represent only \emph{GA} and \emph{greedy} algorithms.

Unsurprisingly, the \emph{GA} algorithm succeeds in discovering an
optimal solutions for small $n$, because a large part of the valid
solution space can be explored, so optimization techniques detect the
best branches. The \emph{greedy} is contrarily sub-optimal. In these
hard configurations, we observe however that this algorithm provides
allocations that fulfill a large majority of peers and are at less
than $15\%$ to the optimal.  Finally, the results of the exact
solutions, especially the impossibility to obtain a perfect
allocation, confirm that non-rival resources impact the overlay
capacity.



\section{Related Works}

The problem of capacity of peer-to-peer networks is a recent and
fairly unexplored topic where related work have focused in main part
on live streaming systems. For instance, \cite{4378424} models the
overlay network as a rooted tree that exhibits capacity constraints on
its links. Determining the maximal overall bandwidth that can be
allocated to peers in such a setting is proved to be NP-hard. This
work could be included in the large existing literature on network
design problems~\cite{netdesign} and resource sharing in networks of
processor-sharing queues~\cite{1011299}. In comparison to these works,
as previously described, the present paper disregards the link
constraints but rather considers that peers have constrained
resources. A similar network model has been considered
in~\cite{streamingcapacity}, where the authors present several
variants of the problem of computing the maximum bandwidth allocation
to all peers in the network. A linear programming approximation is
presented that applies to all instances studied in
\cite{streamingcapacity}, save for the case when the overlay nodes
have bounded outgoing degrees. A similar approach based on primal-dual
algorithms for an edge-weighted network model is studied
in~\cite{cui2004aoc}. We have proposed in this paper a polynomial-time
solution for the optimal allocation, based on a transformation that
maps the problem to a max flow problem on a edge-weighted graph.


Several studies have explored the performance of peer-to-peer systems
for file transfer from one sender to many destinations. A seminal work
in this regard is \cite{Yang2006175}. Most of the other related
studies have focused on analyzing the performance of various data
scheduling strategies, \textit{i.e.}, how long does it take to deliver
a file to $n$ clients in the network. For instance, \cite{QiuS04}
introduces a simple fluid model for analyzing the performance of
Bit-Torrent-like networks. However, the above models neglect the fact
that every peer in the overlay has only a partial view of its
topology. In addition, the peers simultaneously employ data
scheduling, resource allocation, and neighbour management strategies
that is also not taken into account by these models. In contrast, we
consider a snapshot of the peer-to-peer system where every peer
allocates its rival resources to its direct neighbors. Our aim is to
measure the capacity of the network as determined by the peer
neighborhood relationships, \textit{i.e.}, to compute the resource
allocation that actually satisfies the peers' demands.

The problem of resource-driven capacity computation is tighly linked
to the problem of efficient tree construction. It has been shown that
determining a Bounded Degree Spanning Tree (BDST) where no vertex
should have more than $m$ children is however an NP-complete problem
for any degree $m \geq 2$~\cite{garey}.  The BDST is a special case of
$1$-DCDA problem when $c(u)=m, \forall u \in V$. Many related studies
consider determining a spanning tree having the minimum cost on a
weighted graph~\cite{goemans2006mbd}. Interesting variations of this
problem feature non-uniform degree bounds~\cite{KonemannR05} or aim at
minimizing the depth of the spanning tree~\cite{helmick:dlt}. Our
formulation of the $1$-DCDA problem differs in two ways. First, we
consider an unweighted graph as in our model the upload capacities of
the peers act as bottlenecks in the system. In contrast, the above
min-cost optimization problems have been motivated by dimensioning and
reducing the cost of the core network managed by network
operators. Second, these earlier works on spanning trees aim at
spanning \emph{all} nodes in the network while optimizing an objective
function. Differently, the $K$-DCDA problem aims at maximizing the
number of spanned nodes under a node degree constraint. The only
related work in this aspect is~\cite{BlumB05} that studies minimum
trees spanning at least $k$ vertices again in a weighted graph.

When a network is given as a graph with edges associated with weights
and nodes associated with profits, one can formulate a resource
optimization problem such that the profits of the connected nodes
minus the costs of the edges involved is maximized. This is typically
an instance of the Price-Collecting Steiner Tree Problem
(PCSTP)~\cite{Johnson00,Stefan06}, which generalizes the Steiner Tree
Problem. Our problem with one data asset is similar to PCSTP in the
following sense: $1$-DCDA aims to maximize the number of nodes
included in the tree which is equivalent to the case that maximizes
the profit of the nodes when they are associated with a common profit
function and the weights on the edges are zero. However, the problems
are different in that we put constraints on the out degree of the
nodes as otherwise the problem becomes unconstrained.

Finally, numerous works have addressed the design of algorithms aiming
to build peer-to-peer application-layer multicast protocols
(see~\cite{hosseini2007sal} for a survey).  The goal is again to span
all nodes in the overlay, however the optimization objectives here are
application related (\textit{e.g.}, to have a distributed
implementation, to reduce the control message overhead or to ensure a
fast recovery in case of failures). Several related algorithms have
been proposed and extensively analyzed through simulations (see
\textit{e.g.},~\cite{FahmyK07} for a comprehensive study). The most
well-known works include ZigZag and Nice~\cite{TranHD04} that organize
the peer into clusters in order to reduce the control overhead of the
multicast tree. Similarly, TAG~\cite{KwonF05} takes into account the
topology of the underlying network when constructing the multicast
tree in order to reduce its delay.

\section{Conclusions}

This work is a theoretical groundwork for the study of overlay
capacity. We describe an original model and a series of fundamental
results, including a polynomial-time exact algorithm for stationary
regime and a NP-completeness proof with non-rival resources. As the
complexity of this latter problem requires further investigations, we
also describe in this paper two additional contributions: a quite
attractive Bender's decomposition for quick exact solutions and an
efficient heuristic whose experimental performances have proved to be
good. Besides, we raise in this paper various open problems,
\textit{e.g.}, bounded-degree resource allocation in stationary
regime, or the management of dynamic overlays. From a theoretical
point of view, much efforts should be employed to study the $K$-DCDA:
designing approximate algorithms, determining families of overlays on
top of which optimal solutions can be found, analyzing thoroughly
models, \textit{etc.} From an applicative perspective, we would like
to study more deeply efficient bandwidth allocation for improving the
delivery of multiple description video in peer-to-peer streaming
systems. The next steps include the design of distributed
implementations and the study of video-related quality of services.

\bibliographystyle{plain}
\bibliography{IEEEabrv,related}

\end{document}